\documentstyle[prl,aps,epsfig,floats]{revtex}
\begin{document}
\draft
\twocolumn[
\hsize\textwidth\columnwidth\hsize\csname@twocolumnfalse\endcsname

\title{Scaling of the Hysteresis Loop in Two-dimensional Solidification} 
\author{K. Budde$^{a}$, I. Lyuksyutov$^{c*}$, H. Pfn\"ur$^{a**}$, 
G. Godzik$^a$ and H.-U. Everts$^{b}$}
\address{$^{a}$ 
Institut f\"ur Festk\"orperphysik
and $^{b}$Institut f\"ur Theoretische Physik, Universit\"at Hannover,\\
Appelstra\ss e 2, D-30167 Hannover, Germany\\
$^{c}$ 
Department of Physics, Texas A\&M University,\\
College Station, TX 77843-4242, USA 
}
\date{\today}
\maketitle
\begin{abstract} The first order phase transitions 
between a two-dimensional (2d) gas and the 
2d solid of the first monolayer have been studied for the noble gases
Ar, Kr and Xe on a NaCl(100) surface in 
quasi-equilibrium with the three-dimensional (3d) 
gas phase. 
Using linear temperature ramps, 
we show that the widths of the hysteresis 
loops of these transitions as a function of the 
heating rate, $r$,  scales with 
a power law $\propto r^\alpha$ with $\alpha$ between 0.4 
and 0.5 depending on the system.
The hysteresis loops 
for different heating rates
are similar. The island area of the condensed layer was found to grow 
initially with 
a $t^4$ time dependence.
These results are in agreement with  theory, 
which predicts $\alpha = 0.5$ 
and hysteresis loop similarity.
\end{abstract}
%\narrowtext
]

Adsorption isotherm measurements are long known to give 
detailed information about adsorbated-substrate interactions as well as 
about lateral interactions, if combined with structural information. They 
were first introduced into 
surface studies by Langmuir \cite{langmuir}. 
In most cases, adsorption is a first order
phase transition \cite{bruch} with the heat of adsorption as the latent heat, which is expected to 
show the 
typical hysteresis behavior due to overheating and undercooling the adsorbed layer. In this Letter 
we want to concentrate on a specific aspect of hysteresis, scaling of the energy turn-over
as a function of frequency and amplitude, as measured by the hysteresis loop area.

Scaling of the hysteresis loop area by cycling through a first order phase transition has been 
predicted long ago and has been studied in detail in ferromagnetic systems.  
Already in early pioneering work on a 
3D magnet, which dates back to the last century \cite{stein}, first indications
of scaling behavior of the
hysteresis loop  area have been found. 
Recently,  experimental investigations of hysteresis loop scaling
in ultrathin magnetic films 
have been performed \cite{WANG93},\cite{WANG95},\cite{SUEN97}.
However, although power law behavior is found in all studies
\cite{WANG93},\cite{WANG95},\cite{SUEN97}, 
the effective exponents obtained differ 
by typically one order of 
magnitude between theoretical models \cite{theory}
and measurements of energy loss scaling
in ultrathin magnetic films (see discussion in Ref.~\cite{SUEN97}). 
No obvious explanation for
this large discrepancy has come up yet.

Universality mean that scaling of the 
hysteresis loop area to be observable not only 
in ferromagnetic systems, but also, e.g., 
in adsorbed layers, for which, according to our knowledge, 
this property has not been studied in any detail yet. 
This class of systems is characterized not 
only by completely different parameters, 
but also by narrow domain walls of just a few lattice constants,
in contrast to magnetic systems. 
As a result nucleation barriers in adsorbed layers can be much
smaller than in magnetic films.
We have recently predicted scaling behavior of the
adsorption hysteresis loop width, $\mu_{h}$, and of the hysteresis loop area $\cal A$ \cite{Lyuk98}
as a function of amplitude and cycle frequency for the situation close to thermodynamic equilibrium.
A more detailed analysis of the hysteresis behavior
for the case of magnetic film
\cite{Lyuk97},\cite{Lyuk98a} predicts also similarity of different
hysteresis loops.

For our test experiments of scaling of hysteresis loops 
in adsorption of two-dimensional layers, we used the phase transition  to
condensation of a 2d solid of the noble gases Xe, Kr 
and Ar on single crystalline
thin films of NaCl(100) in quasi-equilibrium with the 
3d gas phase at room temperature. 
Due to the van-der-Waals attraction between the noble gas atoms, 
this transition is strongly 
first order, and can be easily measured, as shown below. 
The deviation of the 
chemical potential from the equilibrium value at the phase transition, $\mu$, 
in the case of adsorption plays the role of the 
magnetic field, which was periodically varied 
around the equilibrium value with
amplitude $\mu_0$. 

In our model, the process of adsorption  is
divided into two stages:
nucleation of islands
and growth of these islands. 
The situation when 
the nucleation time $\tau_n \ll \tau$, the cycle time, has been considered in
Ref.~\cite{Lyuk98}.
We call this situation  growth controlled hysteresis. 
For heterogeneous nucleation this condition can be easily fulfilled.

With the additional assumption that an island with 
area ${\cal A} \propto L^2$
grows with a boundary velocity $v(\mu)$ that depends on $\mu$ like
$v = \Gamma \mu $, we obtain for the island of diameter $L$ 
\begin{equation}
\label{lt} 
L(t)- L_0 =
2C{\tau}\Gamma{\mu_0}({\mu(t)\over{4\mu_0}})^2=
{C\Gamma \over 2}r t^2,
\end{equation} 
where $L_0$ is of the order of the size of a stable nucleation center, 
$\mu(t)= rt$ with linear heating rate $r$ (adapted according to the experiments 
carried out), and C a constant of order unity.
We have explicitly used that $({\tau_n/ \tau})^2 \ll 1$.
Eq.\ref{lt} shows that the system behavior is 
controlled by the characteristic  
length $L_{\tau} = {\tau}\Gamma{\mu_0} = 4\mu^2_0\Gamma/r $.
With increase of  $L_{\tau }$ the area of the
adsorption-desorption cycle loop will decrease. 
If $L_N$ is the mean distance between nucleation centers,
then the hysteresis parameters are
defined by the ratio $L_N / L_{\tau}$ \cite{Lyuk97}, \cite{Lyuk98}.

From Eq.\ref{lt} the scaling behavior of
the hysteresis loop width
 $\mu_{h}$ and  of the area of the adsorption-desorption cycle loop $\cal A$
as functions of  $\tau$ and $\mu_0$  
can be evaluated in the limit  $L_N \ll L_{\tau}$, i.e.
when $L_N$ determines the maximum island size.
At  $\mu =\mu_{h}/2$ the typical island size $L\approx L_N$.
The condition  $L_N \ll L_{\tau}$  requires
that $\mu_{h} \ll\mu_0$
and  $L_N \approx  (\Gamma\tau /\mu_0)(\mu_{h})^2$.
If we assume that  $L_N$
does not depend significantly on $\tau$ and  $\mu_0$,
we immediately obtain that 
 $\mu_{h}$ scales as  $\mu_{h} \propto \sqrt{\mu_0 /\tau}\propto \sqrt{r}$.
Under the same conditions of  $L_N$
independent (or weakly dependent) on  $\mu_0$ and  $\tau$,
the hysteresis loops are similar
and scale in the same way as  $\mu_{h}$.
This has been predicted in  \cite{Lyuk97}.
At sufficiently low frequencies, 
the height of hysteresis loop does not 
depend on frequency.  

This part of the frequency dependence is tested in our experiments, 
in which we linearly ramp the surface temperature up and 
down, keeping the ambient gas pressure constant. 
In addition to scaling of the hysteresis loop area, 
the time dependence of island sizes can be 
tested by experiment, which, according to eq.~\ref{lt} 
should grow as $L(t)\propto rt^2$.
As long as the islands don't touch each other, the coverage $\Theta$ depends  
on $L$ as $\Theta\propto L^2\propto r^2t^4$. In a diffraction experiment, the integrated 
intensity of a superstructure spot should be proportional to the 
coverage so that this equation can be tested easily 
measuring the initial time dependence of the 
integrated diffracted intensities.  

Our experiments were carried out in a UHV-chamber 
equipped with high resolution LEED supplemented
by a high resolution optical detector \cite{BUDD98}, 
Auger spectrometer and a quadrupole mass 
spectrometer, at a base pressure of $2 \times 10^{-11}$ mbar. Peak intensities
of LEED diffraction spots were both measured with a Faraday cup and with the optical 
detector, while integrated intensities were obtained from images taken from the optical 
detector using a high resolution slow scan CCD camera. 
The NaCl(100) surfaces were prepared in situ as 3 double layers thick epitaxial films 
by evaporation of NaCl onto a Ge(100) substrate at a surface temperature
of 200 K and subsequent annealing to 550 K. This procedure produces single 
crystalline NaCl layers that are in registry with the Ge substrate on the flat 
terraces and overgrow monoatomic steps of the Ge substrate 
in a carpet-like mode \cite{SCHWENN93}. Therefore the quality of the films is determined
by the step density of the substrate, which had typical terrace length of 400 \AA. The
Ge substrates (size $10\times 10 \times 1$mm) were oriented 
with a precision of 0.1$^\circ$ using a Laue camera and polished 
with diamond pastes and a final chemo-mechanical treatment. These samples were mounted 
onto a He-cooled flow-cryostat. They were heated either by electron bombardment or radiation.
For the measurements only radiation was used. Temperature was measured using a Chromel-Alumel
thermocouple attached to the base plate of the mounting assembly, which was in good thermal 
contact with the Ge substrate. A computerized temperature controller with a nominal resolution of 
0.001 K was used for linear temperature ramps up and down. 
High purity gases were dosed directly onto the sample through a ring-shaped slit 
centered around the surface normal, which could be cooled to 100 K with $\cal {l}$N$_2$. Measurements 
carried out both with gas at room temperature and at 100 K showed that the gas temperature had a 
negligible influence on the results presented below. 
Pressures given below are pressures obtained from a calibration 
with gas exposures from background gas. Homogeneity of the pressure on the surface was also
tested and found to vary negligibly over the surface area seen by the LEED beam.  

\begin{figure}[tb]
\begin{center}
\epsfig{file=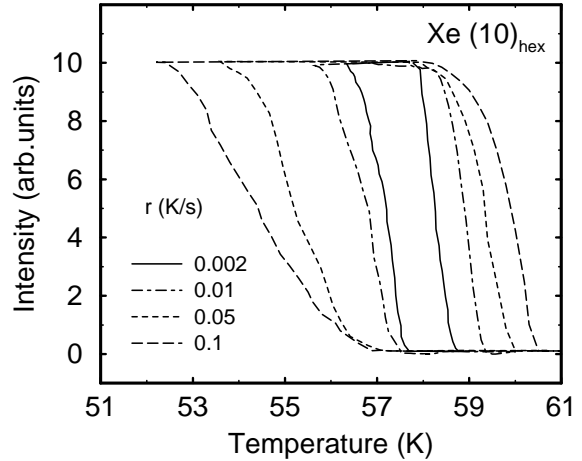,width=0.9\columnwidth}
\end{center}
\caption{Hysteresis loops of the 2d gas-solid phase 
transition of Xe/NaCl(100) at a Xe pressure of $10^{-7}$ mbar. The peak intensity of the
of a first order superstructure spot was monitored 
at the heating rates indicated. \label{xehyst}}
\end{figure}

Xe on NaCl(100) condensed in the first monolayer forms a quasi-hexagonal incommensurate structure 
\cite{SCHWENN94}
so that superstructure diffraction spots were directly used to measure hysteresis  
both with integrated and peak intensities. 
While the first 
monolayer of Ar forms an ordered $(1\times 1)$ structure, a diffuse $(2\times 1)$ structure with 
glide plane symmetry is seen for Kr. Only peak intensities of integral order
beams have been evaluated for the latter two systems. Data of equivalent beams have been averaged 
where available. 

Typical data of the hysteresis during condensation of the quasi-hexagonal Xe layer are shown in 
Fig.~\ref{xehyst}. The linear heating rate was varied by two orders of magnitude between 
0.001 and 0.1 K/s. This rate was limited
at small rates by the resolution of temperature measurement, at high rates by the onset of intermixing 
of second layer condensation. Measurements were taken for gas pressures of $10^{-6}$ and $10^{-7}$ mbar. Not 
surprisingly, slopes during adsorption and desorption are not symmetrical due to the exponential dependence 
of the desorption rate on surface temperature. 
At the highest rate the form during condensation changes due 
to the mentioned onset of intermixing with second layer adsorption, leading also to a decrease of intensity
with decreasing temperature. Inside these limits of heating rate the intensity at saturation did not depend 
on the heating rate. 

\begin{figure}[tb]
\begin{center}
\epsfig{file=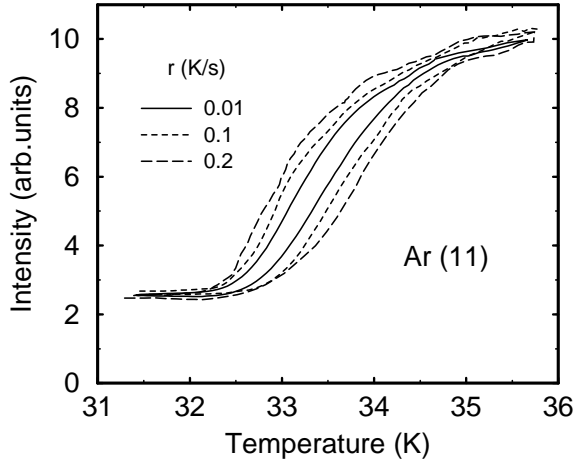,  width=0.9\columnwidth}
\end{center}
\caption{Hysteresis loops of the 2d gas-solid phase transition of Ar/NaCl(100). 
The change of peak intensity
of an first order integral spot was monitored for the 
2d-$(1\times 1)$ Ar solid formed. \label{arhyst}}
\end{figure}

Similar measurements were taken measuring the 
condensation induced changes of  
peak intensities of integral order beams for Ar and Kr on NaCl, which, depending on electron energy of the 
LEED experiment and on diffraction order, can be both positive or negative. An example for Ar induced 
intensity changes is shown in Fig.~\ref{arhyst}. Please note that the dependence on noble gas concentration
need not be linear in these cases. 

\begin{figure}[tb]
\begin{center}
\epsfig{file=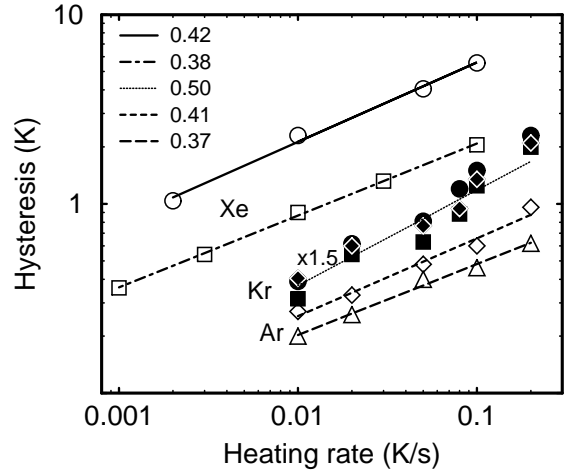, width=0.9\columnwidth}
\end{center}
\caption{Log-log plots of the width of hysteresis loops, $\Delta T$, 
as  a function of the heating rate $r$
in presence of a 3d gas pressure. Xe/NaCl(100): 
$p=1\times 10^{-7}$mbar ($\bigcirc$), 
$p=1\times 10^{-6}$mbar ($\Box$). Full symbols: 
Kr/NaCl(100) at $p=1\times 10^{-7}$mbar (three different orders of diffraction). 
$\diamond$ and $\triangle$: Ar/Nacl(100) 
at $p=1\times 10^{-7}$mbar measured for the (10) and (11) beams,
respectively.\label{loglog}}
\end{figure}

For all three systems, we evaluated the dependence of the widths of the hysteresis loops 
at half maximum intensity on the heating rate $r$. The results are shown as a log-log plot 
in Fig.~\ref{loglog}. Power laws were obtained for all three systems. The effective exponents 
$\alpha$ obtained from these plots are close to 0.4 for Xe and Ar condensation 
(within the statistical uncertainty of about 10\%), 
whereas the average value for Kr of $\alpha = 0.50$ 
actually corresponds exactly to the value expected from the simple model 
of growth controlled hysteresis already mentioned. 
Also the experimental results of Xe and Ar are sufficently close to this value so 
that this model seems to describe the essential physics correctly.

\begin{figure}[tb]
\begin{center}
\epsfig{file=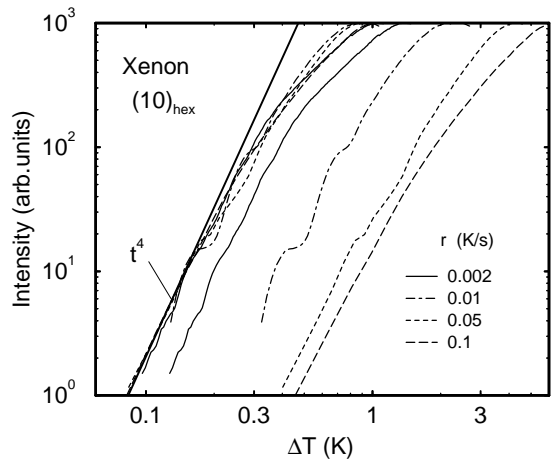, height=0.9\columnwidth, angle=270}
\end{center}
\caption{Test of scaling with time of integral intensities 
of a first order superstructure spot of Xe/NaCl(100),
The curves close to the straight line have been divided by $r^2$.  
\label{timescal}}
\end{figure}

A further test of this model can be carried out by analyzing the time dependence of island growth in 
the initial stages of growth. For this purpose we plotted the integrated intensities of the Xe systems
as a function of time during condensation again on log-log scale (see Fig~\ref{timescal}). 
As mentioned, the model predicts the integrated intensity to 
increase $\propto r^2t^4$, i.e. $\propto (\Delta T)^4/r^2$. Therefore, the 
data in this figure are plotted once as a function 
of $\Delta T$, and a second time scaled by $r^2$. The integrated intensities follow closely the $t^4$
dependence predicted by our simple model, but deviates from it when coverage gets closer to 
saturation. Of course, this plot is very sensitive to the choice 
of time zero, for which we chose the condensation temperature at equilibrium, as estimated from the 
center of the hysteresis curves. This is the earliest possible time. It is fully consistent with the 
assumptions of the model used, which assumes a small nucleation time $\tau_n$. 
The intensity zero was taken 
as the bottom of the hysteresis curves without further adjustments. This result therefore seems to 
nicely corroborate the model assumption of growth controlled hysteresis, which is not limited by 
diffusion on the surface.

These results turned out to be insensitive to additional 
production of anion vacancies on the NACl substrate, 
which is caused by the measuring electron beam. They act as
additional nucleation centers. Though they reduce the maximum size of islands, they obviously
do not change the growth modes, in agreement with the expectations from our model. 
For larger islands, diffusion to the island boundaries of course can no longer be neglected resulting
in deviations from our model. 
Diffusion is particularly important in the second layer on already condensed islands, since there the 
sticking coefficient is much larger than on the bare surface \cite{SCHWENN93}. This process 
increases
the speed of growth, especially at the initial stages.  
Although in a real experiment mass transport 
to the growing boundary by diffusion
is always present, we therefore expect that our main result 
-dynamical scaling- remains valid
for conditions close to equilibrium.

\begin{figure}[tb]
\begin{center}
\epsfig{file=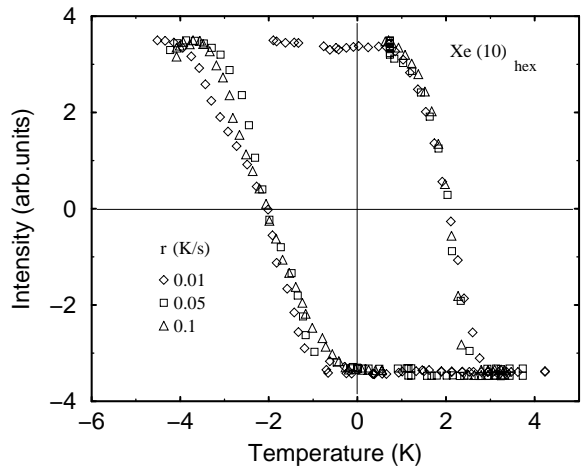, height=0.9\columnwidth, angle=270}
\end{center}
\caption{Hysteresis curves for Xe rescaled by $(\Delta T)^{-\alpha}$. \label{sim}}
\end{figure}

An explicit test of similarity of the hysteresis curves was again carried out for Xe/NaCl(100). 
The rescaled curves (after centering) are shown in Fig.~\ref{sim} for one order of magnitude changes in the heating 
rate. While the evaporation data fit perfectly to a common line, there is more scatter 
in the data during condensation but no general trend for the 
small deviations was found. Therefore, 
also similarity seems to be fulfilled by these data. 

To conclude, we have introduced a new method of
studing adsorption phenomena, observed scaling behavior
for adsorption, found corresponding  critical exponents
and proved similarity of hysteresis loops.
Our studies of scaling 
of the hysteresis  loop areas in 
adsorption gave results in almost quantitative agreement
with theoretical expectations using a simple model. This situation is 
in sharp contrast with the situation in 
magnetic films. Possible reasons are the small width of the interface between islands 
(of the order of one lattice constant)
and/or a low interface energy. The latter would cause small
nucleation barriers. It is consistent 
with the dominant mechanism of growth controlled hysteresis found in our experiments.
For studies of hysteresis phenomena, we have shown that adsorbed films are complementary
to  ultrathin magnetic
films, since they open a completely different parameter space. 
The method we have used can be extended to a broad 
range of 2-D gas - 2-D solid phase transitions. 
In addition this type of measurements can give important information about
dynamics of first order phase transitions on surfaces
and about 2-D interface motion.

We wish to thank 
V.L.Pokrovsky with whom we
extensively discussed the problem of hysteresis loop scaling.   
The work was supported by the Nieders\"achsische Ministerium f\"ur 
Wissenschaft and Kultur, by Volkswagen Stiftung and by the Deutsche Forschungsgemeinschaft.
One of us (I.L.) was partly supported 
by a DOE grant (DE-FG03-96ER 45598).


\begin{references}
\bibitem{theory} References to theoretical work can be found in 
\cite{SUEN97} and \cite{Lyuk98a}.
\end{references}
\end{document}